Reflection and transmission of light at a curved interface: coherent state approach

N.I. Petrov

Lenina Street, 19-39, Istra, Moscow Region 143500, Russia

*Phase-space procedure based on coherent state representation is proposed for investigation of reflection and transmission of light beams at a curved dielectric boundary. Numerical simulations of reflection and transmission of light at various boundaries separating two different dielectrics are carried out. Significant influence of wave-front curvature and polarization of incident beam on the reflectance and transmittance is shown.*

*Introduction:* Micro-optical elements are widely used in modern optical systems, such as light homogenizers, micro-lens arrays, etc. For consideration of large aperture micro-optical systems, conventional methods, such as physical optics, become overly time-consuming. Beam-mode representations do not provide right alternative, since the beam modes cannot be tracked simply through the arbitrarily curved surfaces. Discrete phase-space methods [1, 2] have been proposed as an efficient alternative. These methods represent fields as discrete and finite superpositions of elementary Gaussian beams that can be traced easily in a complicated environment. However, the method usually proposed to discrete fields, the so-called Gabor representation, has been shown numerically unstable [3, 4].

In this paper, the approach based on coherent state representation is proposed for analysis of reflection and transmission of light beam at a curved dielectric surface profile. Coherent states are used in many different areas of physics [5-7]. In [8] the coherent state method was used for consideration of nonparaxial propagation and focusing of wave beams in a graded-index medium. The term "coherent states" was introduced by Glauber [5, 6] for a one-dimensional steady-state quantum oscillator in connection with problems in quantum optics. Such states were constructed and investigated already by Schrodinger in 1926 [9] in order to establish a relationship between the classical and quantum approaches.

*Formulation of the problem:* Consider the curved boundary between two different dielectric media (Fig. 1). For simplicity, the two-dimensional periodically corrugated interface (*y*–independent) with the period $d >> \lambda$ is considered, but the extention to the arbitrary profile of 3D case is straightforward.

The calculation procedure of the reflected and transmitted powers consists of the following steps. At first, the incident beam field is expanded into coherent states, representing elementary Gaussian beams with axis displacement and tilt. Note, that in a sense such Gaussian beams are similar to the complex rays used for simulation of reflection and transmission of beams at a curved interface in [10]. Coherent states (CS) form a full set of functions $\pi^{-1} \int d^2\alpha |\alpha\rangle\langle\alpha| = 1$, so the arbitrary incident field *E(x,0)* can be expanded into a set of CS:

$$E(x,0) = \pi^{-1} \int d^2\alpha \langle x|\alpha\rangle f(\alpha), \tag{1}$$

where $\langle x|\alpha\rangle$ is given by the expression (3), $d^2\alpha = d(\text{Re}\,\alpha)d(\text{Im}\,\alpha)$ is the elementary phase-space volume, $f(\alpha) = \int dx \langle \alpha|x\rangle E(x,0)$ are the amplitudes of the expansion.

Coherent state can be determined as the eigenfunction of the annihilation operator $\hat{a}$:

$$\hat{a}|\alpha\rangle = \alpha|\alpha\rangle, \qquad (2)$$

where $\hat{a} = \dfrac{\hat{x}}{w_0} + i\dfrac{kw_0}{2}\hat{p}_x, \hat{p}_x = -\dfrac{i}{k}\dfrac{\partial}{\partial x}$.

An explicit form of CS is given by a Gaussian beam function

$$|\alpha\rangle = \left(\frac{2}{\pi w_0^2}\right)^{1/4} \exp\left(-\frac{x^2}{w_0^2} + \frac{2x}{w_0}\alpha - \frac{\alpha^2}{2} - \frac{|\alpha|^2}{2}\right), \qquad (3)$$

where the complex eigenvalues $\alpha = \dfrac{x_0}{w_0} + i\dfrac{kw_0}{2}p_{x0} = |\alpha|\exp(i\vartheta)$ determine the initial coordinates $x_0$ of the center of the elementary beam and the angle of its inclination $p_{x0} = n_0\sin\theta_0$ to the z-axis, $n_0$ is the refractive index of medium, $w_0$ is the elementary beam width, $k = 2\pi/\lambda$ is the wavenumber.

Note, in contrast to Fourier-expansion, there is no requirement for orthogonality of functions, owing to CS form overfull function system. The square of the modules $|f(\alpha)|^2 = |\langle\alpha|E\rangle|^2$ determine the incident beam power distribution between the elementary beams (CS). For the incident Gaussian beam $E(x,0) = (2/\pi a_0^2)^{1/4}\exp(-x^2/a_0^2)$ the amplitudes of expansion have the form

$$|f(\alpha)|^2 = \frac{2w_0/a_0}{1+w_0^2/a_0^2}\exp\left\{-|\alpha|^2 + |\alpha|^2\frac{1-w_0^2/a_0^2}{1+w_0^2/a_0^2}\cos 2\vartheta\right\}. \qquad (4)$$

The Gaussian elementary beams (CS) pass through the interface as determined by the corresponding Fresnel coefficients that vary according to the angle of incidence. Usually the Fresnel formulae are known from the plane–wave limit. However, these formulae can be used also for localized wave beams with the beam waists $w > \lambda$ [11, 12].

Finally, total reflected and transmitted powers are determined by a summation of powers of all elementary beams. The reflectance and transmittance are defined as the ratios of reflected and transmitted powers to the incident power, accordingly:

$$r = \frac{P^r}{P^i} = \frac{\int d^2\alpha |f(\alpha)|^2 R(\theta_1^i)}{\int d^2\alpha |f(\alpha)|^2}, \quad t = \frac{P^t}{P^i} = \frac{\int d^2\alpha |f(\alpha)|^2 T(\theta_1^i)}{\int d^2\alpha |f(\alpha)|^2}, \qquad (5)$$

where $R(\theta_1^i)$ and $T(\theta_1^i)$ are the reflection and transmission coefficients, $\theta_1^i$ is the incident angle. The reflection and transmission coefficients for *TE* and *TM* linearly polarized incident beams, accordingly, are given by the expressions

$$R_E = \frac{\left(n_1 \cos\theta_1^i - \sqrt{n_2^2 - n_1^2 \sin^2\theta_1^i}\right)^2}{\left(n_1 \cos\theta_1^i + \sqrt{n_2^2 - n_1^2 \sin^2\theta_1^i}\right)^2}, \quad T_E = \frac{4n_1 \cos\theta_1^i \sqrt{n_2^2 - n_1^2 \sin^2\theta_1^i}}{\left(n_1 \cos\theta_1^i + \sqrt{n_2^2 - n_1^2 \sin^2\theta_1^i}\right)^2};$$

$$R_H = \frac{\left(n_2 \cos\theta_1^i - \frac{n_1}{n_2}\sqrt{n_2^2 - n_1^2 \sin^2\theta_1^i}\right)^2}{\left(n_1 \cos\theta_1^i + \frac{n_1}{n_2}\sqrt{n_2^2 - n_1^2 \sin^2\theta_1^i}\right)^2}, \quad T_H = \frac{4n_1 \cos\theta_1^i \sqrt{n_2^2 - n_1^2 \sin^2\theta_1^i}}{\left(n_2 \cos\theta_1^i + \frac{n_1}{n_2}\sqrt{n_2^2 - n_1^2 \sin^2\theta_1^i}\right)^2}, \qquad (6)$$

where $n_1$ and $n_2$ are the refractive indexes of media 1 and 2, accordingly.

*Results:* For the surface-relief profile $z=s(x)$, the ray (CS) with initial coordinates $(x_0,\theta_0)$ strikes the interface at the point $(x_1, z_1)$, where the corresponding incidence angle $\theta_1^i$ is uniquely determined. For example, the incident angle can be expressed as $\theta_1^i = \varphi - \theta_0$, where $\varphi = \arctan[s'(x)]$, $s'(x) = \frac{ds(x)}{dx}$ is the derivative of the surface profile function with respect to the *x* coordinate.

Results of simulation are presented for different surface profiles, wavefront curvature radiuses and polarizations of incident beam. The parameters for incident beam, surface-relief, and elementary beam are in the ratio $a_0 > d \gg w_0 > \lambda$, where $d$ is the diameter of the single element in corrugated surface (Fig.1). Fig. 2 and Fig. 3 show the reflectance and transmittance as function of sag $h$ of the surface relief $s(x) = h_0 \cos^2(\pi x/d)$ with the period $d=50\mu m$ for TE and TM polarized beams with different wavefront curvature radiuses. Analogical dependences are obtained for the parabolic surface profile $s(x) = s_0 - \frac{(x - x_{0m})^2}{2R_{sc}}$, where $s_0 = \frac{R_L^2}{2R_{sc}}$, $R_L = d/2$ and $R_{sc}$ are the radius and curvature radius of the single element, $x_{0m}$ is the center coordinate of the single element (not shown). Lesser sensitivity of the reflection and transmission to the sag $h$ and wavefront curvature radius $R_f$ changes is observed for parabolic surface profile. Reflectance increases and transmittance decreases with the increase of sag $h$. The reflectance and transmittance are sensitive to the wavefront curvature radius and polarization of the incident beam. For lower-higher index interface the reflectance is lower and transmittance is higher for TM polarized beam if $h < d$. For higher-lower index interface there is no evident difference between TE and TM polarizations. It follows from the simulations that the decrease of the reflectance and increase of the transmittance take place with the increase of the sag $h$ for TM polarization owing to Brewster angle effect.

The proposed method is an efficient alternative to model aperture functions and flat-topped laser beams. Unlike the superposition of off-axis Gaussian function components used in [13], CS decomposition represents the superposition of linearly shifted and spatially rotated beams, forming the full set of functions. Transfer-matrix method [14] can be used to calculate the propagation of off-axis Gaussian beams in optical systems with tilted, displaced and curved optical elements.

*Conclusion:* The CS approach can be used for simulation of intensity distribution of light diffracted by the micro-lens array [15, 16]. As illustrated by simulations, the optical efficiency (transmittance) of such systems strongly depends on the aspect ratio *h/d*. For *h/d* <1 the transmittance of micro-lens arrays exceeds 90%, which is in good agreement with the existing experimental data. The proposed method is an efficient alternative to model aperture functions and flat-topped laser beams.

Thus, the CS based continuous decomposition has been presented for simulation of light beam reflection and transmission at arbitrarily curved surfaces. Reflectance and transmittance of light beam at different surface-reliefs with depths $h \gg \lambda$ are investigated. This method provides effective algorithm to model complicated structures (dielectric antennas, lenses, multiple interfaces, etc.) with the help of combination of wave propagation and ray-tracing procedures.

**Figure captions**

**Fig. 1** Geometrical configuration and coordinate system for a boundary.

**Fig. 2** Reflectance $r$ (solid curves) and transmittance $t$ (dashed curves) versus depth $h$ at lower/higher index interface for different values of incident wavefront curvature radiuses: left - TE polarized beam, right – TM polarized beam; curves 1, $R_f$ = 1500 μm; curves 2, plane wavefront.

**Fig. 3** Reflectance $r$ (solid curves) and transmittance $t$ (dashed curves) versus depth $h$ at higher/lower index interface for different values of incident wavefront curvature radiuses: left - TE polarized beam, right – TM polarized beam; curves 1, $R_f$ = 1500 μm; curves 2, plane wavefront.

Figure 1

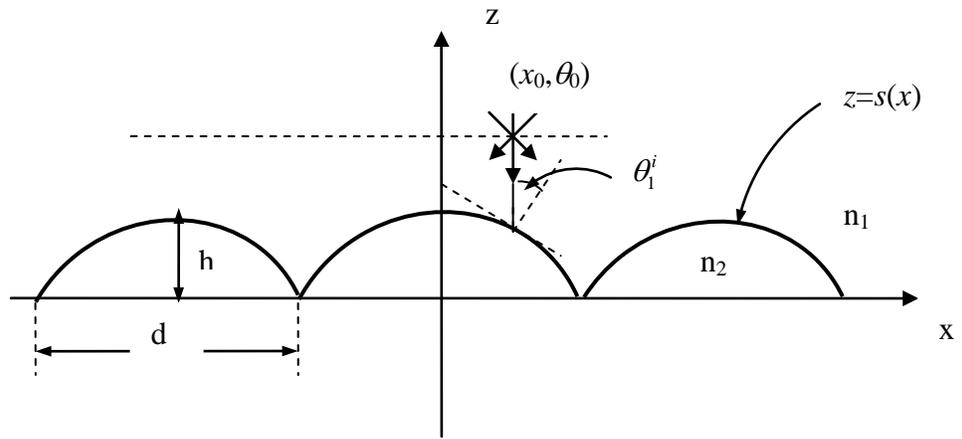

Figure 2

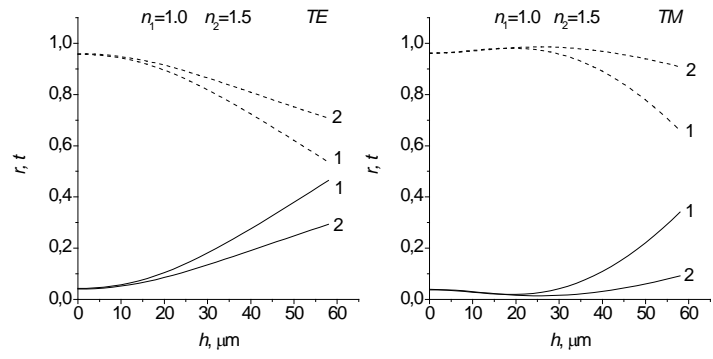

Figure 3

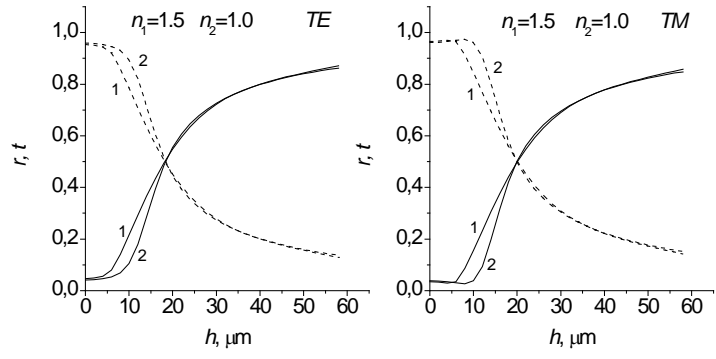